\documentclass[12pt]{article}
\usepackage{amsmath}
\usepackage{latexsym}
\usepackage{amssymb}
\usepackage{fullpage}
\usepackage{tikz}
 \usepackage{colortbl}
\usepackage{epsfig}
\usepackage{graphics}
\usepackage{fancyhdr}

\def\dimo{\noindent\mbox{\sc proof: }}
\def\eproof{\rm\hspace*{\fill}$\Box$\vspace{10pt}}
\newtheorem{defin}{\bf Definition}[section]
\newtheorem{theo}[defin]{Theorem}

\newtheorem{corol}[defin]{Corollary}
\newtheorem{prop}[defin]{Proposition}

\def\cP{\mbox{\boldmath ${\cal P}$}}

\def\meglio {{\quad {\underset {\sim}  \succ}}}

\def\cck{\mbox{\footnotesize \boldmath $k$}}

\def\ccck{\mbox{\boldmath $k$}}

\def\ccq{\mbox{\footnotesize \boldmath $q$}}
\def\cccq{\mbox{\boldmath $q$}}

\def\cccx{\mbox{\boldmath $x$}}

\def\cccy{\mbox{\boldmath $y$}}
\def\cccfi{\mbox{\boldmath $\Phi$}}
\def\cccf{\mbox{\boldmath $F$}}
\def\ccct{\mbox{\boldmath $T$}}

\title{{\sc {Geometry of anonymous
binary social choices that are strategy-proof
}}}

\author{Achille Basile\thanks{Corresponding author.}\\
{ Dipartimento di Scienze Economiche e Statistiche} \\
{Universit\`a Federico II,
80126 Napoli, Italy}\\
{ E-mail: basile@unina.it}
\\
 Surekha Rao \\
{ School of Business and Economics}\\
{ Indiana University Northwest,
Gary, IN 46408}\\
{ E-mail: skrao@iun.edu}\\
 and\\
 K. P. S. Bhaskara Rao\\
{ Department of Computer Information Systems}\\
{Indiana 
University Northwest, Gary, IN 46408}\\
{ E-mail: bkoppart@iun.edu}}

\begin{document}
\maketitle

\thispagestyle{empty}
 
 \newpage
\begin{abstract}
{\normalsize Let $V$ be society whose members express preferences about two alternatives, indifference included.  Identifying anonymous binary social choice functions with  binary functions $f=f(k,m)$ defined over  the integer triangular grid $G=\{(k,m)\in \mathbb{N}_0\times\mathbb{N}_0 : k+m\le |V|\} $, we show that every strategy-proof, anonymous  social choice function  can be described geometrically by  listing, in a sequential manner, groups of segments of G, of equal (maximum possible) length, alternately horizontal and vertical, representative of preference profiles that determine the collective choice of one of the two alternatives. Indeed, we show that every  function which is anonymous and strategy-proof   can be  described in terms of a sequence of  nonnegative integers $(q_1, q_2, \cdots, q_s)$ corresponding to the cardinalities of the mentioned groups of segments. We also analyze the connections between  our present representation with another of our earlier representations involving sequences of majority quotas. 

A Python code is available with the authors for the implementation of any such social choice function.}
\end{abstract}

\bigskip 
{\normalsize AMS Subject Classification: 91B14


}

\thispagestyle{empty}

 \newpage
 
  \setcounter{page}{1}
 
\section{Introduction}
\lhead{\sc \scriptsize Geometry of ...}
\rhead{\sc  \scriptsize Introduction  }
 
 Let $V$ be a finite society that has to implement one of two projects, say alternative $a$ or alternative $b$, based upon the profile $P=(P_v)_{v\in V}$ of preferences expressed by its members. The choice $P_v$ of individual $v$  can be either in favor of $a$, or in favor of $b$, or it can be indifference between the two alternatives\footnote{We can write, for short, $P_v=a, b, ab$, to mean $a\underset{P_v}\succ b, \,b\underset{P_v}\succ a$,\, or\,  $a\underset{P_v}\sim b$ respectively.}. To determine the collective choice corresponding to the profile $P$ of preferences, the society needs to apply a rule, or  {\it social choice function} (scf, for short), we may say. Let  $\phi:\cP\to \{a, b\}$ be such a rule.  Notice that we denote by $\cP$ the set of all possible profiles.

To ensure a fair consideration of the opinions of all agents,  and to prevent strategical false declarations, typical assumptions on the scf $\phi$ are {\it anonymity} and {\it strategy-proofnes} (see Definition \ref{anonsp}).

\bigskip

This paper gives a new geometric characterization of anonymous, strategy-proof scfs, and a corresponding new representation theorem for such functions.

\bigskip
If $\phi$ is an anonymous scf, the collective choice $\phi(P)$ only depends upon the pair $(k(P), m(P))$ of voters that prefer $a$ and $b$, respectively. Consequently, $\phi$ can be seen as a function $f$ from the 
integer triangular grid:
$G=\{(k,m)\in \mathbb{N}_0\times\mathbb{N}_0 : k+m\le |V|\}$, to the set $ \{a, b\}$. Precisely:
$$\phi(P)=f(k(P),m(P))$$
is the law that gives the identification. 

We show that  $\phi$ is in addition strategy-proof if and only if the function $f$ satisfies the following geometric property:
{
\it whenever a point $(k^*,m^*)$ of  $G$  belongs to the inverse image $\{f=a\}$ of the alternative $a$, then all points belonging to the intersection of  $G$ with the cone $A$ in the picture below \footnote{The cone $A$, of vertex $(k^*,m^*)$, is white with dashed axis in Figure $\alpha$, whereas $G$ is in gray.}, also belong to  $\{f=a\}$, namely they represent profiles that determine $a$ as the social choice.}
$$
\begin{tikzpicture}[xscale=0.9, yscale=0.9]
\draw [<->] (0,5.4) -- (0,0) -- (6,0); 
\draw  (0,5)  -- (5,0); 
\node [below right] at (6,0) {\footnotesize$k$};
\node [left] at (0,5.4) {\footnotesize$m$};
\draw [fill=lightgray] (5,0) -- (0,5) --  (0,0);
\draw [fill=white] (1.8,1.2) -- (3.8, 1.2)  -- (5,0) -- (1.8,0);
\draw[fill] (1.8,1.2) circle [radius=0.055]; 
\draw [dashed] [->] (1.8,1.2)  -- (7,1.2);
\draw   (1.8,1.2)  -- (1.8, 0);
\draw [dashed] [->] (1.8,0)  -- (1.8, -1);
\node [above] at (1.8,1.2) {\footnotesize$(k^*,m^*)$}; 
\node [right] at (1.8,0.5) {\scriptsize profiles $\!\in\!\{f=a\}$}; 
\node [right] at (6,0.5) {\footnotesize cone $A$};
\node at (3,-1.1) [below] {\scriptsize Figure $\alpha$};

\draw [<->] (10,5.4) -- (10,0) -- (16,0); 
\draw  (10,5)  -- (15,0); 
\node [below right] at (16,0) {\footnotesize$k$};
\node [left] at (10,5.4) {\footnotesize$m$};
\draw [fill=lightgray] (15,0) -- (10,5) --  (10,0);
\draw [fill=white] (11.8,1.2) -- (11.8, 3.2)  -- (10,5) -- (10,1.2);
\draw[fill] (11.8,1.2) circle [radius=0.055]; 
\draw [dashed] [->] (11.8,1.2)  -- (8.5,1.2);
\draw   (11.8,1.2)  -- (10, 1.2);
\draw [dashed] [->] (11.8,1.2)  -- (11.8, 4.8);
\node [below] at (11.8,1.2) {\footnotesize$(k^*,m^*)$}; 
\node [right] at (10,3) {\scriptsize profiles $\in$ }; 
\node [right] at (10,2.5) {\scriptsize  $\!\in\!\{f=b\}$ };
\node [right] at (6,0.5) {\footnotesize cone $A$};
\node at (13,-1.1) [below] {\scriptsize Figure $\beta$};

\end{tikzpicture}
$$

 The above property  can be formalized as the {\it comprehensiveness} of the inverse image of the alternative $a$. Comprehensiveness is  a property well known for its use in the theory of Core Allocations (see Aliprantis et al. \cite{ABB}). 
 Strategy-profness of an anonymous $f$ is at the same time equivalent to the fact that   the inverse image of the alternative $b$ is comprehensive with respect to the cone with vertex  $(k^*,m^*)$ opposite to $A$ (see Figure $\beta$ above).

\bigskip

As a consequence of the above geometric characterization, we have the following representation theorem.

\begin{theo}\label{main}
 {\it  The anonymous, strategy-proof scfs are all and only the functions $f:G\to \{a, b\}$
that correspond to  finite sequences $\cccq=(q_1, q_2, \dots)$ of integers, with  $q_1\in \{0, 1, \dots, |V|\}$, every further term $q_i\in$
$\{1, 2, \dots, |V|\}$, and $\sum_i\, q_i=|V|+1, $  such that, in order:
\begin{itemize}
\item [$q_1$]  horizontal  segments  in $G$ of maximum possible length, represent profiles  determining, according to $f$, the social choice $a$;
\item [$q_2$] vertical segments in $G$ of maximum possible length given the above, represent profiles  determining, according to $f$, the social choice $b$;
\item [$q_3$] horizontal segments in $G$ of maximum possible length given the above, represent profiles  determining, according to $f$, the social choice $a$;
\item [$q_4$] vertical segments in $G$ of maximum possible length given the above, represent profiles  determining, according to $f$, the social choice $b$;
\end{itemize}
and so on, until $G$ is filled.
}\end{theo}

The figure below illustrates the representation theorem. 
$$\begin{tikzpicture}[xscale=0.3, yscale=0.3]
\draw[help lines] (0,0) grid (20,20);
\draw (0,20) --(20,0);  \draw [ultra thick] [magenta] (0,0) --(20,0); \draw [ultra thick] [magenta] (0,1) --(19,1); \draw  [ultra thick] [magenta] (0,2) --(18,2); 
\draw  [ultra thick] [magenta] (0,3) --(17,3); 
\draw  [ultra thick] [magenta] (0,4) --(16,4); 
\draw (0,20) --(20,0);  \draw [ultra thick] [blue] (0,5) --(0,20); \draw [ultra thick] [blue] (1,5) --(1,19); \draw  [ultra thick] [blue] (2,5) --(2,18);
\draw (0,20) --(20,0);
\draw  [ultra thick] [magenta] (3,5) --(15,5); \draw  [ultra thick] [magenta] (3,6) --(14,6); 
\draw (0,20) --(20,0);  \draw [ultra thick] [blue] (3,7) --(3,17); \draw [ultra thick] [blue] (4,7) --(4,16); \draw  [ultra thick] [blue] (5,7) --(5,15); \draw  [ultra thick] [blue] (6,7) --(6,14); \draw  [ultra thick] [blue] (7,7) --(7,13); \draw  [ultra thick] [blue] (8,7) --(8,12); 
\draw  [ultra thick] [magenta] (9,7) --(13,7); 
\draw (0,20) --(20,0);  \draw [ultra thick] [blue] (9,8) --(9,11); \draw [ultra thick] [blue] (10,8) --(10,10); \draw  [ultra thick] [blue] (11,8) -- (11,9); \draw  [ultra thick] [blue] (12,8) -- (12,8);
\draw[fill] [ultra thick] [blue] (12,8) circle [radius=0.055];
 \node at (10,-1) [below] {\scriptsize Figure 0};

 \draw[help lines] (24,0) grid (44,20);
\draw (24,20) --(44,0);  \draw [ultra thick] [blue] (24,0) --(24,20); \draw [ultra thick] [blue] (25,0) --(25,19); \draw  [ultra thick] [blue] (26,0) --(26,18); 
\draw  [ultra thick] [magenta] (27,0) --(44,0); \draw  [ultra thick] [magenta] (27,1) --(43,1); 
\draw (24,20) --(44,0);  \draw [ultra thick] [blue] (27,2) --(27,17); \draw [ultra thick] [blue] (28,2) --(28,16); \draw  [ultra thick] [blue] (29,2) --(29,15); \draw  [ultra thick] [blue] (30,2) --(30,14); 
\draw  [ultra thick] [magenta] (31,2) --(42,2); \draw  [ultra thick] [magenta] (31,3) --(41,3); 
\draw  [ultra thick] [magenta] (31,4) --(40,4); \draw  [ultra thick] [magenta] (31,5) --(39,5); 
\draw  [ultra thick] [magenta] (31,6) --(38,6); 
\draw (24,20) --(44,0);  \draw [ultra thick] [blue] (31,7) --(31,13);
\draw  [ultra thick] [magenta] (32,7) --(37,7); 
\draw  [ultra thick] [magenta] (32,8) --(36,8); 
\draw  [ultra thick] [magenta] (32,9) --(35,9); \draw  [ultra thick] [magenta] (32,10) --(34,10); 
\draw  [ultra thick] [magenta] (32,11) --(33,11); 
\draw[fill] [ultra thick] [magenta] (32,12) circle [radius=0.055];
 \node at (34,-1) [below] {\scriptsize Figure $0_{bis}$ };

\end{tikzpicture}
$$

In both figures we have $|V|=20$. Blue vertical lines represent profiles corresponding to which the social choice is $b$; the magenta horizontal lines  represent profiles corresponding to which the social choice is $b$. In Figure 0 we have $\cccq=(5, 3, 2, 6, 1, 4)$ and the scf represented implements the alternative $a$ if all voters are indifferent between $a$ and $b$. 
In Figure $0_{bis}$ we have
 $\cccq=(0, 3, 2, 4, 5, 1, 6 )$ and the scf represented implements the alternative $b$ if all voters are indifferent. 
 
 Given the role of the integers $q_i$, we give the name {\it $\{a,b\}$-list} to the sequence $\cccq$.

\bigskip
{\bf Comparison  with previous literature.}  Despite  the general circumstance according to which representation formulas are a quite obvious topic to investigate, to the best of our knowledge, only recently representation theorems of anonymous strategy-proof scfs in presence of indifference have been proposed. If indifference is not allowed,   \cite[Corollary of page 63]{M} completely describes the $|V|+2$ existing anonymous strategy proof scfs. Allowing for indifference, in the literature we find   \cite[Theorem 2] {LP},
and  the representation \cite[Theorem 2.7]{KPS}. Anonymous strategy-proof scfs are represented in \cite{KPS} by means of suitably monotone sequences ({\it up and down sequences}) $\ccck$ of majority quotas, in this way providing a natural extension  of \cite[Corollary of page 63]{M} for representing the $2^{|V|+1}$  anonymous strategy-proof scfs  that exist when agents can declare indifference.

\cite[Theorem 2] {LP}, \cite[Theorem 2.7]{KPS} and Theorem \ref{main}, are definitely three different results. The first two having been compared in \cite{KPS},  in Section \ref{quattro} we compare in details the last two, also providing  the conversion formulas (Theorem \ref{confronto} and its reverse) to move back and forth from the present geometric representation to the extended quota majority representation \cite[Theorem 2.7]{KPS}.

\bigskip


The remaining sections of the paper are organized as follows. In Section \ref{due}, we set the notations and introduce dually monotone functions.  By means of such functions  the strategy-proofness of anonymous scfs is characterized and  also related to the comprehensiveness of the inverse image of the alternatives. Corollary \ref{diciotto} is an intermediate representation theorem.
Section \ref{dually} is devoted to the characterization of dual monotonicity my means of $\{a, b\}$-lists. Our main result Theorem \ref{main} just combines Corollary \ref{diciotto} with Theorem \ref{dual monotone}. The last section presents the proofs of Theorems \ref{dual monotone} and \ref{confronto}.

A Python code, written by the authors, that implements a given anonymous strategy-proof scf and shows the geometric representation is available. This code can also be used to convert proper majority quotas into the corresponding $\{a, b\}$-list and viceversa.

 \section{Basic notions, notation}\label{due}
 \lhead{\sc \scriptsize Geometry of ...}
\rhead{\sc  \scriptsize Basic notions, notation  }

 Throughout the sequel we assume that the cardinality of $V$ is $n$. Given a profile $P$, the set $D(a,P)$  (resp. $D(b,P)$) denotes  the subset of $V$ consisting of voters that prefer $a$ over $b$ (resp. $b$ over $a$). The agents that are indifferent are obviously those of $V\setminus [D(a,P)\cup D(b,P)]$. We also set
 $k(P)=|D(a,P)|$ and $m(P)=|D(b,P)|$.
 
 Typically a scf, i.e. a function that maps profiles of preferences to alternatives that have to be implemented as the result of a collective choice, will be denoted by $\phi$. 
 The following definitions are well established and widely employed in the literature after the pioneering works \cite{G},\cite{S}; see also \cite{BBM} for specific reference to the binary case.
 
 \begin{defin}\label{anonsp}
A scf $\phi$ is:
\begin{itemize}
\item[] {\bf anonymous}, if \,\, $\phi(P)=\phi(P\circ \sigma)=\phi(\, (\,P_{\sigma(v)}\,)_{v\in V}\,),$ for every profile $P$ and for every permutation $\sigma$ of $V$.
\item[] {\bf non-manipulable}, if \,\, $ \phi(P_v, P_{-v}) \meglio_{P_v} \phi(Q_v, P_{-v})$, for every voter $v$, for every profile $P$, and for every weak ordering $Q_v$.
\end{itemize}
\end{defin}

 This paper  deals with {\it binary scfs}, i.e. only two alternatives are considered. For the sake of brevity we write scf to mean binary scf.
 
 Let us denote by $\cccfi$ the set of all anonymous scfs $\phi$.

Let us denote by $\cccf$ the set of all functions $f=f(k,m)$ defined on the integer triangular grid:

$$G=\{(k,m)\in [0,n]^2: k+m\le n\},$$ and with values $ f(k,m)\in \{a, b\}.$
\,\, We set $\ell:=n-k-m$.

\begin{defin}
Two profiles $P$ and $Q$ are said to be equivalent when $k(P)=k(Q)$ and $m(P)=m(Q)$. In this case we shall write $P\equiv Q$.
\end{defin}

\begin{defin}
For every $(k,m)\in G$, by $P(k,m)$ we denote the profile where the first $\ell$ agents are indifferent between $a$ and $b$, the subsequent $k$ agents declare to prefer $a$ and the last $m$ agents declare to prefer $b$.
\end{defin}

The proof of the next two propositions are trivial.

\begin{prop}
The quotient $\cP\!/\!\!\equiv$ and the grid $G$ are in a one-to-one correspondence by means of the map
$$P\!/\!\!\equiv \,\mapsto (k(P), m(P)), \mbox{ whose inverse is } (k,m)\mapsto P(k,m)/\!\!\equiv.$$
Moreover, $$P\equiv Q \Rightarrow \phi(P)=\phi(Q), \forall \phi\in \cccfi.$$
\end{prop}

\begin{prop}
The function sets $\cccfi$ and $\cccf$ are in a one-to-one correspondence by means of the map
$$ L: \phi\in\cccfi  \mapsto L(\phi)\in \cccf, \mbox{ where  $L(\phi)$ is the function  $f_\phi$ defined by }  f_\phi(k,m)= \phi(P(k,m)).$$
The inverse of $L$ is  $$f\in \cccf \mapsto \phi_f\in \cccfi \mbox{ where } \phi_f(P):=f(k(P), m(P)).$$
\end{prop}

We now introduce the concept of dual monotonicity for functions $f(k,m)$.

\begin{defin}
We say that $f\in \cccf$ is {\bf dually monotone} if the following implication holds true:

$$f(k,m)=a \Rightarrow f(k+1,m)=f(k,m-1)=a.$$
\end{defin}

As we see below, dual monotonicity can be defined also with reference to the alternative $b$ instead of $a$. We denote by $C_a$ and $C_b$ the two cones in the $km$-plane defined respectively by 
$$ (\alpha,\beta)\in C_a \Leftrightarrow \alpha\ge0 \,\,\&\,\, \beta\le0,$$ 
$$ (\alpha,\beta)\in C_b \Leftrightarrow \alpha\le0 \,\,\&\,\, \beta\ge0.$$

If we adopt the notation $H+C$, when $H\subseteq G$ and $C$ is one of the two above cones, for the set
$\{(k',m')\in G: (k',m')=(k,m)+(\alpha, \beta), {\rm \,\,for \,\, some \,\,} (k,m)\in H {\rm \,\,and \,\, some\,\, } (\alpha, \beta)\in C\},$ then we easily see that:
\begin{prop}
For $f\in \cccf$, TFAE:
\begin{enumerate}

\item $f$ is {\bf dually monotone} 

\item $f(k,m)=b \Rightarrow f(k-1,m)=f(k,m+1)=b$

\item $\{f=a\}+ C_a\subseteq \{f=a\}$
\item $\{f=b\}+ C_b\subseteq \{f=b\}$
\end{enumerate}
\end{prop}

Property 3.  above says  that the set $\{f=a\}$ is
comprehensive. Equivalently the set $\{f=b\}$ is
comprehensive. Notice that , however, the cones are different.


It is obvious from the previous proposition that if $f$ is dually monotone, then $f(0,n)=a$ corresponds to the constant function $f=a$, and $f(n,0)=b$ corresponds to the constant function $f=b$.

\begin{theo}
$\phi\in\cccfi$ is strategy-proof if and only if $f_\phi$ is dually monotone.

\end{theo}
\dimo
We write here $f$ instead of $f_\phi$, for short.

Let us assume that $\phi$ is anonymous and strategy-proof.  If $f$ is not dually monotone, we have $f(k,m)=a$ and either $f(k+1,m)=b$ or  $f(k,m-1)=b$. In both cases we get a contradiction. We analyze the case $f(k+1,m)=b$, the other case being analogous. By definition we have $\phi(P(k,m))=a$ and  $\phi(P(k+1,m))=b$. The difference between the profiles $P(k,m)$ and $P(k+1,m)$ is: one agent that in $P(k,m)$ is indifferent, under $P(k+1,m)$ prefers $a$. So, this agent can advantageously manipulate $P(k+1,m)$ by declaring indifference.

For the converse let us suppose that $f_\phi$ is dually monotone whereas $\phi$ is manipulable. So, we have an agent $v_0$, a profile $P$, and a profile $Q=(Q_{v_0}, P_{-v_0})$ such that
$$(*) \qquad \phi(Q_{v_0}, P_{-v_0})\underset{P_{v_0}}\succ \phi(P_{v_0}, P_{-v_0}).$$
Let us set $|D(a,P)|=k, |D(b,P)|=m$ and $|D(a,Q)|=k', |D(b,Q)|=m'$.
The relation $(*)$ can be written as

$$(*) \qquad f(k',m') \underset{P_{v_0}}\succ f(k,m),$$
and only two cases are possible from $(*)$. 

\bigskip
Case $v_0$ prefers $a$: It means that $f(k',m')=a$. Comparing $(Q_{v_0}, P_{-v_0})$ and $(P_{v_0}, P_{-v_0})$, we see that if the agent $v_0$ $Q_{v_0}$-prefers $b$, then $m'=m+1$ and $k'=k-1$.
If the agent $v_0$ is $Q_{v_0}$-indifferent between $a$ and $b$, then $m'=m$ and $k'=k-1$.

In both cases the  dual monotonicity of $f$ gives us that $f(k,m)=a$ and this is impossible.

\bigskip
Case $v_0$ prefers $b$: It is analogous.
\eproof

\begin{corol}\label{diciotto}
A scf $\phi$ is anonymous and strategy-proof if and only if  it is of the form $\phi_f$ for a dually monotone $f\in \cccf$.
\end{corol}

\section{Dually monotone functions on $G$}\label{dually}
 \lhead{\sc \scriptsize Geometry of ...}
\rhead{\sc  \scriptsize Dually monotone functions  }

In this section we give a representation of all dually monotone functions. Because of Corollary \ref{diciotto}, this will give us the representation of all anonymous, strategy-proof scfs that have been stated as Theorem \ref{main}.

 Let $\cccq$ be a finite sequence $(q_1, q_2, \dots)$. We say that it is an {\bf $\{a,b\}$-list} if $q_1\in \{0, 1, \dots, n\}$, every further term $q_i$ belongs to
$\{1, 2, \dots, n\}$, and $\sum_i\, q_i=n+1.$ Given $\cccq$, we  define the sets
$$Q_a=\{(0,q_1-1), (q_2, q_1+q_3-1), (q_2+q_4, q_1+q_3+q_5-1), \dots  \},$$ and

$$Q_b=\{(q_2-1,q_1), (q_2+q_4-1, q_1+q_3), (q_2+q_4+q_6-1, q_1+q_3+q_5), \dots  \}.$$
It is elementary to check that $$[Q_a+ C_a]\cap [Q_b+ C_b]=\O,$$
hence the following $f_{\ccq}$ is a well defined function on $G$.

$$(k,m)\in G\mapsto f_{\ccq} (k,m)=\left\{ 
\begin{array} {ll}
a,   & \mbox{ if  } (k,m)\in Q_a+ C_a  \\

b,   & \mbox{ if  } (k,m)\in Q_b+ C_b.\\
\end{array}
\right.$$
The structure of the function $f_{\ccq}$ has been already described more intuitively  in the statement of Theorem \ref{main}. Indeed, it is easy to check that $f_{\ccq}$  can be described as follows.

In order:
\begin{itemize}
\item [$q_1$]  horizontal  segments  in $G$ of maximum possible length are mapped by $f_{\ccq}$ to $a$;
\item [$q_2$] vertical segments in $G$ of maximum possible length given what above, are mapped by $f_{\ccq}$ to $b$;
\item [$q_3$] horizontal segments in $G$ of maximum possible length given what above, are mapped by $f_{\ccq}$ to $a$;
\item [$q_4$] vertical segments in $G$ of maximum possible length given what above, are mapped by $f_{\ccq}$ to $b$;
\end{itemize}
and so on.

\begin{theo}\label{dual monotone}
An element $f\in\cccf$ is dually monotone if and only if for some { $\{a,b\}$-list} $\cccq$ one has $f=f_{	\ccq}.$
\end{theo}

The proof of the theorem will be given in the Appendix.

\bigskip
We conclude the section rephrasing Theorem \ref{main}  as follows.

{\bf Theorem \ref{main}}  {\it A scf $\phi$ is anonymous and strategy-proof if and only if $\phi=\phi_{f_{\ccq}}$, for some { $\{a,b\}$-list} $\cccq$.}

\section{Comparison with proper extended quota majority methods}\label{quattro}
 \lhead{\sc \scriptsize Geometry of ...}
\rhead{\sc  \scriptsize Comparison with proper extended quota majority methods  }

In \cite{KPS} we proved that the scfs that are anonymous and strategy-proof can be characterized as proper extended quota majority methods (denoted by $\phi_{\cck}$; see \cite[Definition 2.2]{KPS} for the definition of the scf $\phi_{\cck}$), the majority quotas $\ccck=(k_0, k_1, \dots, k_r)$ forming a sequence satisfying  the
following
 {\bf up and down} conditions.
 
\bigskip
down-up:
$$
\begin{tabular}{ccc}
\hline
\multicolumn{1}{|c}{$0$} & \multicolumn{1}{|c}{$%
< k_{r-1}<...<k_{5}<k_{3}<k_{1}<k_{0}<k_{2}<k_{4}<k_{6}<...<$} & \multicolumn{1}{|c|}{%
$n+1=k_{r}$} \\ \hline
&  &  \\ \hline
\multicolumn{1}{|c}{$0=k_{r}$} & \multicolumn{1}{|c}{$%
<...<k_{5}<k_{3}<k_{1}<k_{0}<k_{2}<k_{4}<k_{6}<...<k_{r-1}<$} & \multicolumn{1}{|c|}{%
$n+1$} \\ \hline
\end{tabular}
$$

up-down:
$$
\begin{tabular}{ccc}
\hline
\multicolumn{1}{|c}{$n+1=k_{r}$} & \multicolumn{1}{|c}{$%
>...>k_{5}>k_{3}>k_{1}>k_{0}>k_{2}>k_{4}>k_{6}>...>k_{r-1}>$} & \multicolumn{1}{|c|}{%
$0$} \\ \hline
\multicolumn{1}{l}{} & \multicolumn{1}{l}{} & \multicolumn{1}{l}{} \\ \hline
\multicolumn{1}{|c}{$n+1$} & \multicolumn{1}{|c}{$%
>k_{r-1}>...>k_{5}>k_{3}>k_{1}>k_{0}>k_{2}>k_{4}>k_{6}>...>$} & \multicolumn{1}{|c|}{%
$0=k_{r}$} \\ \hline
\end{tabular}
$$
At the same time, as we have seen in Theorem \ref{main}, the scfs under consideration are identifiable by means of $\{a, b\}$-lists $\cccq$ as scfs $\phi=\phi_{f_{\ccq}}$. It is therefore natural to investigate formulas that connect up and down majority quotas  $\ccck$ and $\{a, b\}$-lists $\cccq$ when they identify the same scf, namely when
$\phi_{\cck}=\phi_{f_{\ccq}}.$

The purpose of the present section is to provide a one-to-one map $T$ that transforms an up and down sequence $\ccck$ into an $\{a,b\}$-list $\cccq$ such that the scf represented is the same. 

It will be a map like 
$$(q_1, q_2, \dots, q_s)=T(k_0,k_1,\dots,k_r)$$
i.e. we shall  give  formulas like

$$q_i=T_i(k_0,k_1,\dots,k_r), \qquad i\le s=s(r).$$

\begin{theo}\label{confronto}
Let $\phi$ be an anonymous, strategy-proof scf and $\ccck=(k_0,k_1,\dots,k_r)$ the corresponding up and down sequence of majority quotas (i.e. $\phi=\phi_{\ccck}$).
\begin{enumerate}
\item If $\phi$ selects the alternative $a$ when the collectivity is unanimously indifferent between the two alternatives $a$ and $b$\footnote{We recall that, by the definition of extended quota majority method, $k_r=0$ (resp. $k_r=n+1$) corresponds to scfs choosing the alternative $a$ (resp. $b$) for a collectivity which is unanimously indifferent between the two alternatives $a$ and $b$.}, the $\{a,b\}$-list satisfying the equation
$\phi_{\cck}=\phi_{f_{\ccq}}$ is $(q_1, q_2, \dots, q_r, q_{r+1})$, where
$$
q_1=k_{r-1}^\circ,
\, q_2=k_{r-2},\,
q_3=k_{r-3}^\circ-k_{r-1}^\circ,\,
q_4=k_{r-4}-k_{r-2},\,\dots $$ 
$$ 
q_i=\left\{ 
\begin{array} {ll}
k_{r-i}^\circ-k_{r-i+2}^\circ,   & \mbox{ if $5\le i\le r$ is odd }   \\

k_{r-i}-k_{r-i+2},   & \mbox{ if $5\le i\le r$ is even },\\
\end{array}
\right. 
q_{r+1} =n+1-\sum_1^r  q_i
$$

\item If $\phi$ selects the alternative $b$ when the collectivity is unanimously indifferent between the two alternatives $a$ and $b$, the $\{a,b\}$-list satisfying the equation
$\phi_{\cck}=\phi_{f_{\ccq}}$ is $(0, q_{1+1}, \dots,  q_{1+r}, q_{r+2})$, where
$$
q_{1+1}=k_{r-1},
\, q_{1+2}=k_{r-2}^\circ,\,
q_{1+3}=k_{r-3}-k_{r-1},\,
q_{1+4}=k_{r-4}^\circ-k_{r-2}^\circ,\,\dots $$ 
$$ 
q_{1+i}=\left\{ 
\begin{array} {ll}
k_{r-i}-k_{r-i+2},   & \mbox{ if $5\le i\le r$ is odd }   \\

k_{r-i}^\circ-k_{r-i+2}^\circ,   & \mbox{ if $5\le i\le r$ is even },\\
\end{array}
\right. 
q_{r+2} =n+1-\sum_1^r  q_{1+i}
$$
\end{enumerate}

\end{theo}
The proof of the theorem will be given in the Appendix.

\bigskip
Above formulas can be both written as  $\cccx=\ccct \cccy$ if $\ccct$ is the lower triangular matrix of order $r$:

$$\ccct={\small \left[ 
\begin{array}{cccccccccccccc}
1 & 0 & 0 & 0 & 0 & 0 & 0 & 0 & 0 & 0 & 0 & 0 & 0 & 0 \\ 
0 & 1 & 0 & 0 & 0 & 0 & 0 & 0 & 0 & 0 & 0 & 0 & 0 & 0 \\ 
-1 & 0 & 1 & 0 & 0 & 0 & 0 & 0 & 0 & 0 & 0 & 0 & 0 & 0 \\ 
0 & -1 & 0 & 1 & 0 & 0 & 0 & 0 & 0 & 0 & 0 & 0 & 0 & 0 \\ 
0 & 0 & -1 & 0 & 1 & 0 & 0 & 0 & 0 & 0 & 0 & 0 & 0 & 0 \\ 
0 & 0 & 0 & -1 & 0 & 1 & 0 & 0 & 0 & 0 & 0 & 0 & 0 & 0 \\ 
&  &  &  &  &  &  &  &  &  &  &  &  &  \\ 
&  &  &  &  & ... & ... & ... &  &  &  &  &  &  \\ 
&  &  &  &  &  &  &  &  &  &  &  &  &  \\ 
0 & 0 & 0 & 0 & 0 & 0 & 0 & 0 & 0 & -1 & 0 & 1 & 0 & 0 \\ 
0 & 0 & 0 & 0 & 0 & 0 & 0 & 0 & 0 & 0 & -1 & 0 & 1 & 0 \\ 
0 & 0 & 0 & 0 & 0 & 0 & 0 & 0 & 0 & 0 & 0 & -1 & 0 & 1%
\end{array}%
\right] }$$

and, for formula 1. of Theorem \ref{confronto}, $$\cccx '=(q_1, q_2,\dots, q_r), \mbox{ and } \cccy '=(k^\circ_{r-1}, k_{r-2},k^\circ_{r-3}, \dots, k^\circ_{r-(r-1)}, k_{r-r}).$$

For 2. of Theorem \ref{confronto}, we have to consider $$\cccx '=(q_{1+1}, q_{1+2},\dots, q_{1+r}), \mbox{ and } \cccy '=(k_{r-1}, k_{r-2}^\circ,k_{r-3}, \dots, k_{r-(r-1)}, k_{r-r}^\circ).$$

By using the inverse of $\ccct$

$${\ccct}^{-1}=\left[ 
\begin{array}{cccccccccccccc}
1 & 0 & 0 & 0 & 0 & 0 & 0 & 0 & 0 & 0 & 0 & 0 & 0 & 0 \\ 
0 & 1 & 0 & 0 & 0 & 0 & 0 & 0 & 0 & 0 & 0 & 0 & 0 & 0 \\ 
1 & 0 & 1 & 0 & 0 & 0 & 0 & 0 & 0 & 0 & 0 & 0 & 0 & 0 \\ 
0 & 1 & 0 & 1 & 0 & 0 & 0 & 0 & 0 & 0 & 0 & 0 & 0 & 0 \\ 
1 & 0 & 1 & 0 & 1 & 0 & 0 & 0 & 0 & 0 & 0 & 0 & 0 & 0 \\ 
0 & 1 & 0 & 1 & 0 & 1 & 0 & 0 & 0 & 0 & 0 & 0 & 0 & 0 \\ 
&  &  &  &  &  &  &  &  &  &  &  &  &  \\ 
&  &  &  &  & ... & ... & ... &  &  &  &  &  &  \\ 
&  &  &  &  &  &  &  &  &  &  &  &  &  \\ 
0 & 1 & 0 & 1 & 0 & 1 & 0 & 1 & 0 & 1 & 0 & 1 & 0 & 0 \\ 
1 & 0 & 1 & 0 & 1 & 0 & 1 & 0 & 1 & 0 & 1 & 0 & 1 & 0 \\ 
0 & 1 & 0 & 1 & 0 & 1 & 0 & 1 & 0 & 1 & 0 & 1 & 0 & 1%
\end{array}%
\allowbreak \right] $$  
we solve, with respect to $\ccck$, given $\cccq$, the equation $\phi_{\cck}=\phi_{f_{\ccq}}$ as follows.

\bigskip
{\bf Theorem \ref{confronto} reversed}
{\it Let $\phi$ be an anonymous, strategy-proof scf and $\cccq=(q_1,\dots,q_s)$ the corresponding $\{a,b\}$-list (i.e. $\phi=\phi_{f_{\ccq}}$).
\begin{enumerate}
\item If $\phi$ selects the alternative $a$ when the collectivity is unanimously indifferent between the two alternatives $a$ and $b$\footnote{Hence $q_1>0$.}, the  sequence of majority quotas satisfying the equation
$\phi_{\cck}=\phi_{f_{\ccq}}$ is $(k_0, k_1, \dots, k_{r})$, where $r=s-1$,
$$(k^\circ_{r-1}, k_{r-2},k^\circ_{r-3}, \dots, k^\circ_{r-(r-1)}, k_{r-r})' ={\ccct}^{-1}(q_1,\dots,q_r)'
\mbox{ and } k_{r} =0
$$

\item If $\phi$ selects the alternative $b$ when the collectivity is unanimously indifferent between the two alternatives $a$ and $b$\footnote{Hence $\cccq=(0,q_{1+1},\dots,q_{1+r}, q_{r+2})$.}, the  sequence of majority quotas satisfying the equation
$\phi_{\cck}=\phi_{f_{\ccq}}$ is $(k_0, k_1, \dots, k_{r})$, where $r=s-2$,
$$(k_{r-1}, k_{r-2}^\circ,k_{r-3}, \dots, k_{r-(r-1)}, k^\circ_{r-r})' ={\ccct}^{-1}(q_{1+1},\dots,q_{1+r})'
\mbox{ and } k_{r} =n+1
$$
\end{enumerate}}
Explicitly, we have 
\begin{itemize}
\item for  case 1. \newline
$k_{r-1}^\circ=q_1$, \, $k_{r-2}=q_2$,\, $k_{r-3}^\circ=q_1+q_3$,\, $k_{r-4}=q_2+q_4$,\, $k_{r-5}^\circ=q_1+q_3+q_5$,\, $k_{r-6}=q_2+q_4+q_6$, ...

\item for case 2.\newline
$k_{r-1}=q_{1+1}$, \, $k_{r-2}^\circ=q_{1+2}$,\, $k_{r-3}=q_{1+1}+q_{1+3}$,\, $k_{r-4}^\circ=q_{1+2}+q_{1+4}$,\, $k_{r-5}=q_{1+1}+q_{1+3}+q_{1+5}$,\, $k_{r-6}^\circ=q_{1+2}+q_{1+4}+q_{1+6}$, ...
\end{itemize}

\section{Appendix: proofs}
\lhead{\sc \scriptsize Geometry of ...}
\rhead{\sc  \scriptsize Proofs  }

\subsection{Proof of Theorem \ref{dual monotone}}
Suppose $f$ is given and it is dually monotone. 
Let $q_1$ be the cardinality of the set $$\{m\in [0,n]: f(0,m)=a\}.$$  In case $q_1=n+1$, this means $f(0,n)=a$ and because of dual monotonicity, $f=f_{(q_1)}$.

Suppose  instead that $q_1<n+1$. Since $f$ is dually monotone, 
$$q_1-1=\max\{m\in [0,n]: f(0,m)=a\}\footnote{The elements of the set $\{m\in [0,n]: f(0,m)=a\}$, and of similar sets  introduced later,  are necessarily consecutive.}
$$
hence we have $f(0,m)=a,\,\, \forall m\in[0, q_1[$ and $f(0,q_1)=b$, and by dual monotonicity
$f(0,q_1-1)+C_a \subseteq \{f=a\}$. The following figure, where magenta lines indicate points of $G$ that $f$ maps to $a$, illustrates the situation.

 $$\begin{tikzpicture}[xscale=0.3, yscale=0.3]
\draw[help lines] (0,0) grid (20,20);
\draw (0,20) --(20,0);  \draw [ultra thick] [magenta] (0,0) --(20,0); \draw [ultra thick] [magenta] (0,1) --(19,1); \draw  [ultra thick] [magenta] (0,2) --(18,2); \draw  [ultra thick] [magenta] (0,3) --(17,3); \node at (0,3) {\!\!\!\!\!\!\!\!\!\!\!\!\!\!\!\!\!\!\!\!\!\!\!\!\!\!\!\!\!\!\!\!\!{\scriptsize $f(0, q_1-1)=a$}};
\node at (0,4) {\!\!\!\!\!\!\!\!\!\!\!\!\!\!\!\!\!\!\!\!\!\!\!\!\!{\scriptsize $f(0, q_1)=b$}};
\node at (10,-1) [below] {\scriptsize Figure 1};
\end{tikzpicture}
$$
 Now, let $q_2$ be the cardinality of the set $$\{k\in [0,n-q_1]: f(k,q_1)=b\},$$ or, in other words, let $q_2-1$ be the maximum of such set.  In case $q_2-1=n-q_1$, we have 
  $(n-q_1, q_1)+C_b\subseteq \{f=b\}$ since $f$ is dually monotone. Hence $f=f_{(q_1,q_2)}.$
 
 \bigskip
 If instead we have $q_2-1<n-q_1$, then $f(k,q_1)=b,\,\,\forall k\in [0, q_2[$, and $f(q_2,q_1)=a.$
 Moreover
 $(q_2-1, q_1)+C_b\subseteq \{f=b\}$ since $f$ is dually monotone. 
 Figure 2 extends the previous one, and illustrates the situation. The blue lines indicate points of $G$ that $f$ maps to $b$
  
$$\begin{tikzpicture}[xscale=0.3, yscale=0.3]
\draw[help lines] (0,0) grid (20,20);
\draw (0,20) --(20,0);  \draw [ultra thick] [magenta] (0,0) --(20,0); \draw [ultra thick] [magenta] (0,1) --(19,1); \draw  [ultra thick] [magenta] (0,2) --(18,2); \draw  [ultra thick] [magenta] (0,3) --(17,3); \node at (0,3) {\!\!\!\!\!\!\!\!\!\!\!\!\!\!\!\!\!\!\!\!\!\!\!\!\!\!\!\!\!\!\!\!\!{\scriptsize $f(0, q_1-1)=a$}};
\node at (0,4) {\!\!\!\!\!\!\!\!\!\!\!\!\!\!\!\!\!\!\!\!\!\!\!\!\!{\scriptsize $f(0, q_1)=b$}};
\draw [ultra thick] [blue] (0,4) --(0,20); 
\draw [ultra thick] [blue] (1,4) --(1,19);
\draw [ultra thick] [blue] (2,4) --(2,18);
\draw [ultra thick] [blue] (3,4) --(3,17);
\draw [ultra thick] [blue] (4,4) --(4,16);
\draw [ultra thick] [blue] (5,4) --(5,15);\node at (5,4) {\,\,\,\,\,\,\,\,\,\,\,\,\,\,\,\,\,\,\,\,\,\,\,\,\,\,\,\,\,\,\,{\scriptsize $f(q_2-1, q_1)=b$}};
\node at (10,-1) [below] {\scriptsize Figure 2};
\end{tikzpicture}
$$
Let $q_3$ be the cardinality of the set $$\{m\in [q_1,n-q_2]: f(q_2,m)=a\},$$ or, in other words, let $q_1+q_3-1$ be the maximum of such set. Again we have two cases. If $q_1+q_3-1=n-q_2$, then
 $(q_2, n-q_2)+C_a\subseteq \{f=a\},$ hence $f=f_{(q_1,q_2,q_3)}$.
 
 \bigskip 
 If $q_1+q_3-1<n-q_2$, then $f(q_2,m)=a,\,\,\forall m\in[q_1, q_1+q_3-1]$, and $f(q_2, q_1+q_3)=b$. Moreover $(q_2, q_1+q_3-1)+C_a\subseteq \{f=a\}$.
 
 \bigskip
 We illustrate the situation with the next update of Figure 2.

$$\begin{tikzpicture}[xscale=0.3, yscale=0.3]
\draw[help lines] (0,0) grid (20,20);
\draw (0,20) --(20,0);  \draw [ultra thick] [magenta] (0,0) --(20,0); \draw [ultra thick] [magenta] (0,1) --(19,1); \draw  [ultra thick] [magenta] (0,2) --(18,2); \draw  [ultra thick] [magenta] (0,3) --(17,3); \node at (0,3) {\!\!\!\!\!\!\!\!\!\!\!\!\!\!\!\!\!\!\!\!\!\!\!\!\!\!\!\!\!\!\!\!\!{\scriptsize $f(0, q_1-1)=a$}};
\draw [ultra thick] [blue] (0,4) --(0,20); 
\draw [ultra thick] [blue] (1,4) --(1,19);
\draw [ultra thick] [blue] (2,4) --(2,18);
\draw [ultra thick] [blue] (3,4) --(3,17);
\draw [ultra thick] [blue] (4,4) --(4,16);
\draw [ultra thick] [blue] (5,4) --(5,15);\node at (5,4) {$\,\,\,b$};
\draw [ultra thick] [magenta] (6,4) --(16,4); \draw [ultra thick] [magenta] (6,5) --(15,5); \draw  [ultra thick] [magenta] (6,6) --(14,6);  \node at (6,6) {$\!\!\!a$}; \node at (16,6) [right] {\scriptsize $a$ on the left is $f(q_2, q_1+q_3-1)$};
\node at (10,-1) [below] {\scriptsize Figure 3};

\end{tikzpicture}
$$
Let $q_4$ be the cardinality of the set $$\{k\in [q_2,n-q_1-q_3]: f(k,q_1+q_3)=b\},$$ or, in other words, let $q_2+q_4-1$ be the maximum of such set.

Again we have two cases. If $q_2+q_4-1=n-q_1-q_3$, then
 $(n-q_1-q_3,q_1+q_3)+C_b\subseteq \{f=b\},$ hence $f=f_{(q_1,q_2,q_3,q_4)}$.
 
 \bigskip 
 If $q_2+q_4-1<n-q_1-q_3$, then $f(k, q_1+q_3)=b,\,\,\forall k\in[q_2, q_2+q_4-1]$, and $f(q_2+q_4, q_1+q_3)=a$. Moreover $(q_2+q_4-1, q_1+q_3)+C_b\subseteq \{f=b\}$.
 
 \bigskip
 We illustrate the situation with the next update of the previous figures.

 $$\begin{tikzpicture}[xscale=0.3, yscale=0.3]
\draw[help lines] (0,0) grid (20,20);
\draw (0,20) --(20,0);  \draw [ultra thick] [magenta] (0,0) --(20,0); \draw [ultra thick] [magenta] (0,1) --(19,1); \draw  [ultra thick] [magenta] (0,2) --(18,2); \draw  [ultra thick] [magenta] (0,3) --(17,3); \node at (0,3) {\!\!\!\!\!\!\!\!\!\!\!\!\!\!\!\!\!\!\!\!\!\!\!\!\!\!\!\!\!\!\!\!\!{\scriptsize $f(0, q_1-1)=a$}};
\draw [ultra thick] [blue] (0,4) --(0,20); 
\draw [ultra thick] [blue] (1,4) --(1,19);
\draw [ultra thick] [blue] (2,4) --(2,18);
\draw [ultra thick] [blue] (3,4) --(3,17);
\draw [ultra thick] [blue] (4,4) --(4,16);
\draw [ultra thick] [blue] (5,4) --(5,15);\node at (5,4) {$\,\,\,b$};
\draw [ultra thick] [magenta] (6,4) --(16,4); \draw [ultra thick] [magenta] (6,5) --(15,5); \draw  [ultra thick] [magenta] (6,6) --(14,6);  \node at (6,6) {$\!\!\!a$};
\draw [ultra thick] [blue] (6,7) --(6,14);
\draw [ultra thick] [blue] (7,7) --(7,13);
\draw [ultra thick] [blue] (8,7) --(8,12);
\draw [ultra thick] [blue] (9,7) --(9,11);\node at (9,7) {$\,\,\,b$};
\node at (16,6) [right]{\scriptsize $a$ on the left is $f(q_2, q_1+q_3-1)$};
\node at (16,7) [right]{\scriptsize $b$ on the left is $f(q_2+q_4-1, q_1+q_3)$};
\node at (18,4) [right]{\scriptsize $b$ on the left is $f(q_2-1, q_1)$};
\node at (10,-1) [below] {\scriptsize Figure 4};

\end{tikzpicture}
$$
The procedure continues by defining $q_5$ as the cardinality of the set $$\{m\in [q_1+q_3,n-q_2-q_4]: f(q_2+q_4,m)=a\},$$ or, in other words,  $q_1+q_3+q_5-1=\max\{m\in [q_1+q_3,n-q_2-q_4]: f(q_2+q_4,m)=a\}$, and so on.
The procedure stops as soon as the sum of the $q_i$'s  reaches $n+1$. 

\bigskip
That every $f_{\ccq}$ is dually monotone is obvious by definition.
\eproof

\subsection{Proof of Theorem \ref{confronto}}
We show the assertion in 2., assuming that the given sequence of majority quotas is 
$$
\begin{tabular}{ccc}
\hline
\multicolumn{1}{|c}{$0$} & \multicolumn{1}{|c}{$%
<k_{r-1}<...<k_{5}<k_{3}<k_{1}<k_{0}<k_{2}<k_{4}<k_{6}<...<$} & \multicolumn{1}{|c|}{%
$n+1=k_{r}$} \\ \hline
\end{tabular}
$$
in the other cases the argument is quite similar. The straightforward application of the definition of extended quota majority method (see \cite[Definition 2.2]{KPS}), and the fact that for the dual quotas $k_i^\circ$ the above reciprocal ordering is reversed, leads to identify that, with reference to the grid $G$,
\begin{itemize}
\item the profiles for which $\phi$ selects $a$ as social choice are those that  corresponds to points of coordinates $(k,m)$ such that:\newline
$k\ge k_0$, or $(k,m)\in [k_1,k_0[\times [0, k_0^\circ[$, or $(k,m)\in [k_3,k_1[\times [0, k_2^\circ[$, or\newline
$(k,m)\in [k_5,k_3[\times [0, k_4^\circ[$, or, ... , $(k,m)\in [k_{r-1},k_{r-3}[\times [0, k_{r-2}^\circ[$. 
\item the profiles for which $\phi$ selects $b$ as social choice are those that  corresponds to points of coordinates $(k,m)$ such that:\newline
$m\ge k_0^\circ$, or $(k,m)\in [0,k_1[\times [k_2^\circ, k_0^\circ[$, or $(k,m)\in [0, k_3[\times [k_4^\circ, k_2^\circ[$, or\newline
$(k,m)\in [0,k_5[\times [k_6^\circ, k_4^\circ[$, or, ... , $(k,m)\in [0, k_{r-1}[\times [k_r^\circ, k_{r-2}^\circ[$. 
\end{itemize}
Since $k_r^\circ=0$, there are no full horizontal lines identifying profiles for which the social choice is $a$ (in other words $q_1=0$), whereas there are $k_{r-1}$ full vertical lines of profiles for which the social choice is $b$.  So, $q_2=k_{r-1}$. This is illustrated in Figure 5 where $|V|=20, k_r= 21, k_{r-1}=3, k_{r-2}=19, k_{r-3}=7, k_{r-4}=14, \dots$

$$
\begin{tikzpicture}[xscale=0.3, yscale=0.3]
\draw[help lines] (0,0) grid (20,20);
\draw (0,20) --(20,0);  \draw [ultra thick] [blue] (0,0) --(0,20); \draw [ultra thick] [blue] (1,0) --(1,19); \draw  [ultra thick] [blue] (2,0) --(2,18); 
\node at (10,-1) [below] {\scriptsize Figure 5};
\draw[help lines] (24,0) grid (44,20);
\draw (24,20) --(44,0); 
 \draw [ultra thick] [blue] (24,0) --(24,20); \draw [ultra thick] [blue] (25,0) --(25,19); \draw  [ultra thick] [blue] (26,0) -- (26,18); 
 \draw  [ultra thick] [magenta] (27,0) --(44,0); \draw  [ultra thick] [magenta] (27,1) --(43,1); 
 \node at (34,-1) [below] {\scriptsize Figure 6};
\end{tikzpicture}
$$

Now, horizontal lines that represent profiles for which the social choice is $a$ will depart from $k=k_{r-1}$ and there are $k_{r-2}^\circ$ many of such lines, i.e. $q_3=k_{r-2}^\circ$. This is illustrated in Figure 6


Now, further vertical  lines that represent profiles for which the social choice is $b$ will depart from $k=k_{r-2}^\circ$ and there are $k_{r-3}-k_{r-1}$ many of such new lines, i.e. $q_4=k_{r-3}-k_{r-1}$. This is illustrated in Figure 7.

Again, further horizontal  lines that represent profiles for which the social choice is $a$ will depart from $k=k_{r-3}$ and there are $k_{r-4}^\circ-k_{r-2}^\circ$ many of such new lines, i.e. $q_5=k_{r-4}^\circ-k_{r-2}^\circ$. This is illustrated in Figure 8.

$$\begin{tikzpicture}[xscale=0.3, yscale=0.3]
\draw[help lines] (0,0) grid (20,20);
\draw (0,20) --(20,0);  \draw [ultra thick] [blue] (0,0) --(0,20); \draw [ultra thick] [blue] (1,0) --(1,19); \draw  [ultra thick] [blue] (2,0) --(2,18); 
\draw  [ultra thick] [magenta] (3,0) --(20,0); \draw  [ultra thick] [magenta] (3,1) --(19,1); 
\draw (0,20) --(20,0);  \draw [ultra thick] [blue] (3,2) --(3,17); \draw [ultra thick] [blue] (4,2) --(4,16); \draw  [ultra thick] [blue] (5,2) --(5,15); \draw  [ultra thick] [blue] (6,2) --(6,14); 
\draw[help lines] (24,0) grid (44,20);
\draw (24,0) --(44,0);  \draw [ultra thick] [blue] (24,0) --(24,20); \draw [ultra thick] [blue] (25,0) --(25,19); \draw  [ultra thick] [blue] (26,0) --(26,18); 
\draw  [ultra thick] [magenta] (27,0) --(44,0); \draw  [ultra thick] [magenta] (27,1) --(43,1); 
\draw (24,20) --(44,0);  \draw [ultra thick] [blue] (27,2) --(27,17); \draw [ultra thick] [blue] (28,2) --(28,16); \draw  [ultra thick] [blue] (29,2) --(29,15); \draw  [ultra thick] [blue] (30,2) --(30,14); 
\draw  [ultra thick] [magenta] (31,2) --(42,2); \draw  [ultra thick] [magenta] (31,3) --(41,3); 
\draw  [ultra thick] [magenta] (31,4) --(40,4); \draw  [ultra thick] [magenta] (31,5) --(39,5); 
\draw  [ultra thick] [magenta] (31,6) --(38,6); 
 \node at (34,-1) [below] {\scriptsize Figure 8};
  \node at (10,-1) [below] {\scriptsize Figure 7};
\end{tikzpicture}
$$

The argument flows like above, alternating horizontal and vertical lines, till the grid $G$ is completely filled, according to the sequence $\ccck =(\dots, 14,7,19,3,21)\footnote{Observe that Figure $0_{bis}$ is one of the possible completion of Figure 8. Precisely the one corresponding to $\ccck =(8, 14,7,19,3,21)$.} $.
\eproof

\end{document}